\documentclass[aps,pre,floats,twocolumn,showpacs,floatfix]{revtex4-1}
\usepackage{amsmath}                       
\usepackage{amsfonts}
\usepackage{amssymb}
\usepackage{wasysym}
\usepackage{longtable}

\usepackage{graphicx}
\usepackage{graphicx,epsfig}
\usepackage{bm}
\usepackage{xcolor}

\begin{document}
\title{Elementary processes governing the evolution of road networks}

\author{Emanuele Strano$^{1,2}$, Vincenzo Nicosia$^{3,4}$, Vito
  Latora$^{4,5,6}$, Sergio Porta$^{2}$, Marc
  Barthelemy$^{7,8}$}

\affiliation{%
  $^1$~Laboratory of Geographic Information Systems (LASIG), School of
  Architecture, Civil and Environmental Engineering (ENAC), Ecole
  Polytechnique F\'ed\'erale de Lausanne (EPFL), Switzerland}
\affiliation{%
  $^2$~Urban Design Studies Unit, University of Strathclyde, Glasgow,
  United Kingdom}
\affiliation{%
  $^3$~Computer Laboratory, University of Cambridge, Cambridge, United
  Kingdom}
\affiliation{%
  $^4$~Laboratorio sui Sistemi Complessi, Scuola Superiore di Catania, 
  Via San Nullo 5/i, 95123 Catania, Italy}
\affiliation{%
  $^5$School of Mathematical Sciences, Queen Mary, University of
  London, London,  United Kingdom}
\affiliation{
  $^6$~Dipartimento di Fisica e Astronomia, Universit\'a di Catania and
  INFN, Via S. Sofia 64, 95123 Catania, Italy}
\affiliation{%
  $^7$~Institut de Physique Th\'eorique, CEA, CNRS-URA 2306, F-91191,
  Gif-sur-Yvette, France}
\affiliation{%
  $^8$~Centre d'Analyse et de Math\'ematiques Sociales, EHESS, 130, avenue
  de France, 75244 Paris, France}

\begin{abstract}
  Urbanisation is a fundamental phenomenon whose quantitative
  characterisation is still inadequate. We report here the empirical
  analysis of a unique data set regarding almost 200 years of
  evolution of the road network in a large area located north of Milan
  (Italy). We find that urbanisation is characterised by the
  homogenisation of cell shapes, and by the stability throughout time
  of high--centrality roads which constitute the backbone of the urban
  structure, confirming the importance of historical paths. We show
  quantitatively that the growth of the network is governed by two
  elementary processes: (i) `densification', corresponding to an
  increase in the local density of roads around existing urban centres
  and (ii) `exploration', whereby new roads trigger the spatial
  evolution of the urbanisation front. The empirical identification of
  such simple elementary mechanisms suggests the existence of general,
  simple properties of urbanisation and opens new directions for its
  modelling and quantitative description.
\end{abstract}

\maketitle

%%%%%%%%%%%%%%%%%%%%%%%%%%%%%%%%%%%%%%%%%%%%%%%%%%%%%%
%\section*{Introduction}

Urbanisation is a fundamental process in human history~\cite
{Mumford:book}, and is increasingly affecting our
environment~\cite{McDonald:2011} and our
society~\cite{Dye:2008,Glaeser:2011,Barnes:2011}. The fraction of the
world population living in urban areas has recently grown beyond 50\%
and is expected to rapidly increase in the near future~\cite{UN}. The
challenges presented by such a rapidly urbanising world are complex
and difficult to handle. They range from the increasing dependence of
our society on fossil fuels~\cite{Newman:1999}, to the emergence of
socio-spatial inequalities~\cite{UN:slum,UN:2008}, and to the up-rise
of serious environmental issues~\cite{Satterthwaite:book}. Controlling
urbanisation has already proven to be a difficult task, and it will
become even more difficult in the near future, due to the large and
unprecedented magnitude of urban expansion~\cite {Angel:book}, and to
the intrinsic complexity of this phenomenon, which generates cities
that are continuously changing over time~\cite {Hall:1997} in a
non-homogeneous fashion~\cite{Seto:2010}. Trying to understand the
elementary spatial mechanisms that govern urbanisation, leaving out
specific historical, geographical, social and cultural factors, is
nowadays more important than ever, especially because policy makers,
professionals and researchers are actively looking for new paradigms
in urban planning, land management and
ecology~\cite{UN-Habitat:2010,Turner:2007, Ramalho:2011}. The
existence of acultural and non-demographic drivers of urbanisation has
been investigated using different approaches~\cite {Batty:2008},
including classical studies in regional sciences~\cite{Wilson:1970},
theory of complex systems~\cite{Batty:fractal,Batty:book,Makse:1995},
urban theory~\cite{Nikos:book} and remote
sensing~\cite{Hannes:2012}. A consistent amount of literature in urban
history and morphology indicates that roads are a fundamental driver
in urban evolution and, at the same time, one of the long-lasting
constituent elements of urban
forms~\cite{Marshall:book,Southworth:book}. In the last decade,
complex networks theory has provided significant contributions to the
quantitative characterisation of urban street
patterns~\cite{delta_centrality,Xie:2007,Barthelemy:2011}. Several
studies have shown that road networks not only play a central role in
the spatial organisation of urban
areas~\cite{Lammer:2006,Cardillo:2006}, but are important for the
dynamical processes occurring on
them~\cite{Barthelemy:2011,Bettencourt:2007,Bettencourt10,Balcan:2009,Porta:2011}
and for the evolution of urban systems in
general~\cite{Hanson:2004,Xie:2009}. However, a quantitative analysis
of the historical development of urbanisation in metropolitan areas is
still missing, and empirical evidence of the basic mechanisms
governing urbanisation dynamics is still lacking. Our paper addresses
this problem, and provides a study of the evolution over two centuries
of the road network in a large urban area located north of Milan
(Italy).

\begin{figure*}
\centering
\includegraphics [width=7in] {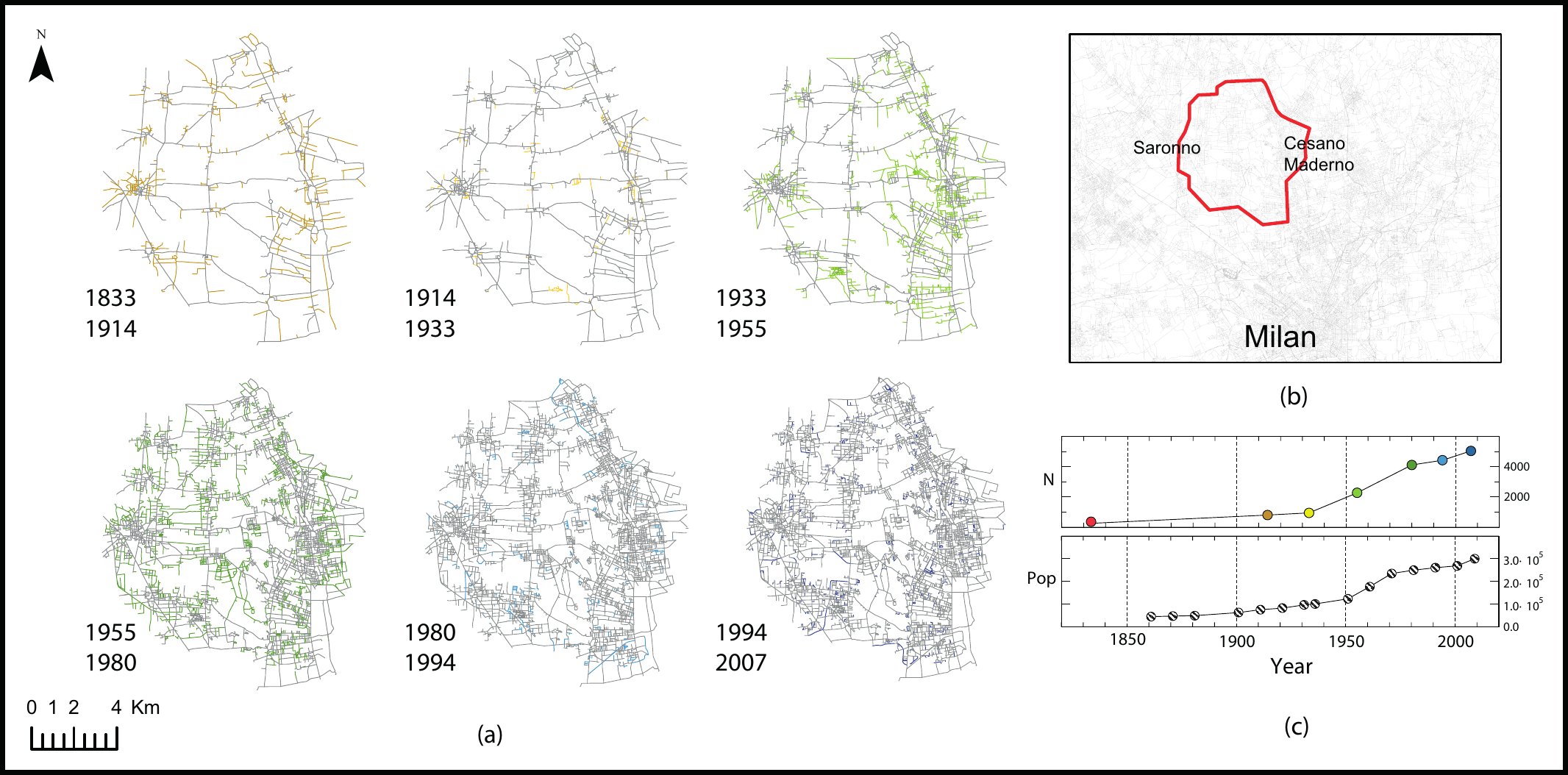}	
  \caption{\textbf{(a)} Evolution of the road network from 1833 to
    2007. For each map we show in grey all the nodes and links already
    existing in the previous snapshot of the network, and in colors
    the new links added in the time window under
    consideration. \textbf{(b)} Map showing the location of the
    Groane area in the metropolitan region of Milan. \textbf{(c)} Time
    evolution of the total number of nodes $N$ in the network and of
    the total population in the area (obtained from census data).}
  \label{fig:panel1}
\end{figure*}

The area under study, known as {\em Groane}, covers a surface of $125
~\text{km}^2$, includes $29$ urban centres within $14$ municipalities,
and has essentially evolved along two main radial paths, connecting
Milan to Como and Milan to Varese, respectively. The former path was
constructed by the Romans during the II century B.C., while the latter
was created during the XVI century. In the last two centuries, the
Groane area has faced a complex process of conurbation, changing from
a polycentric region into a completely urbanised area. This
conurbation process is common to many large European metropolitan
regions~\cite{EU:2006}. Despite local differences, it is possible to
identify four distinct phases that characterize urbanization in the
Groane area: i) {\em Rural phase}, (1800-1918): fundamentally
pre-industrial, fully based on agricultural economy, with no major
transportation infrastructures present. ii) {\em Early-urban phase}
(1918-1945): between world wars period, still significantly based on
agriculture economy, witnessing the first appearance of rail network,
first small-scale sparse industrial colonisation, and limited
expansion of rural settlements around the historical centres. iii)
{\em Urban-industrial phase} (1945-1990): remarkable sprawled
residential and industrial development (especially textile and
mechanics) along with population growth and highway construction. iv)
{\em Metropolitan post-industrial phase} (1990- 2012): decline of
industrial activities, slower-paced urban sprawl, former polycentric
organisation overwhelmed by the metropolitan continuum caused by the
merging of expanded centres, increased long-range mobility due to the
development of high speed trains and large highway systems
\cite{Turri:2000}. It is important to note that this area has never
been subjected to overall large-scale planning efforts, one reason
being that $14$ different administrative bodies preside over $14$
different municipalities.  By importing historical topographical and
photogrammetrical data (for details see the Methods section) into a
Geographical Information System (GIS) environment, we reconstructed
the detailed road system (including minor streets) at seven different
points in time, $t=1,2,\ldots,7$, respectively corresponding to years:
$1833, 1914, 1933, 1955, 1980, 1994, 2007$.  For each time, we
constructed the associated primal graph~\cite{Barthelemy:2011},
i.e. the graph where the nodes represent street junctions and the
links correspond to road segments. The geographical information
sources used to construct the primal graphs are reported in
Table~\ref{table:table1}.  

\begin{table*}
\centering
\footnotesize
\begin{tabular*}{\hsize}{@{\extracolsep{\fill}}rrrr}
Date&Source&Owner&Format\cr
\hline
1833&  Topographical Map of Lombardy-Venetia Kingdom & Italian Military Geographic Institute & Raster\cr
1914& Map of Italy &Italian Military Geographic Institute &Raster\cr
1933&Map of Italy&Italian Military Geographic Institute &Raster\cr
1955&Aerial Photography Survey &Italian Military Geographic Institute &Raster\cr
1980&Lombardy Regional Map &Lombardy Region& Raster\cr
1994&Lombardy Regional Map&Lombardy Region&Raster\cr
2007&Mosaic of Urban Municipalities Plans&Lombardy Region&Vectorial\cr
\hline
\end{tabular*}
\caption{List of geographical information sources used to construct
  the data set.}
\label{table:table1}
\end{table*}

In Fig.~\ref{fig:panel1}a we display the road networks at different
times, showing how the initial small separate villages have grown by
the addition of new nodes and links, eventually merging together in an
homogeneous pattern of streets. We denote by $G_t \equiv G(V_t, E_t)$
the graph at time $t$, where $V_t$ and $E_t$ are respectively the set
of nodes and the set of links at time $t$.  The number of nodes at
time $t$ is then $N(t)= |V_t|$, while $E(t) = |E_t|$ is the number of
links.  By definition, we have $V_t = V_{t-1} \cup \Delta V_t$ and
$E_t = E_{t-1} \cup \Delta E_t$, where $\Delta V_t$ and $\Delta E_t$
are, respectively, the set of new nodes and the set of new links added
to the system in the time window $]t-1,t]$. In the following we will
    study the structure of the graph $G_t$ at different times $t$,
    focusing in particular on the properties of the new links. As we
    will show, a quantitative analysis of the temporal evolution of
    the system will be able to reveal in particular the existence of
    two simple mechanisms that drive the evolution of the road network
    over time.

%%%%%%%%%%%%%%%%%%%%%%%%%%%%%%%%%%%%%%%%%%%%%%%%%%%%%%
\section*{Results}

\subsection*{Characterising the growth of the road network.} 
\begin{figure*}
\centering 
\includegraphics[width=7in]{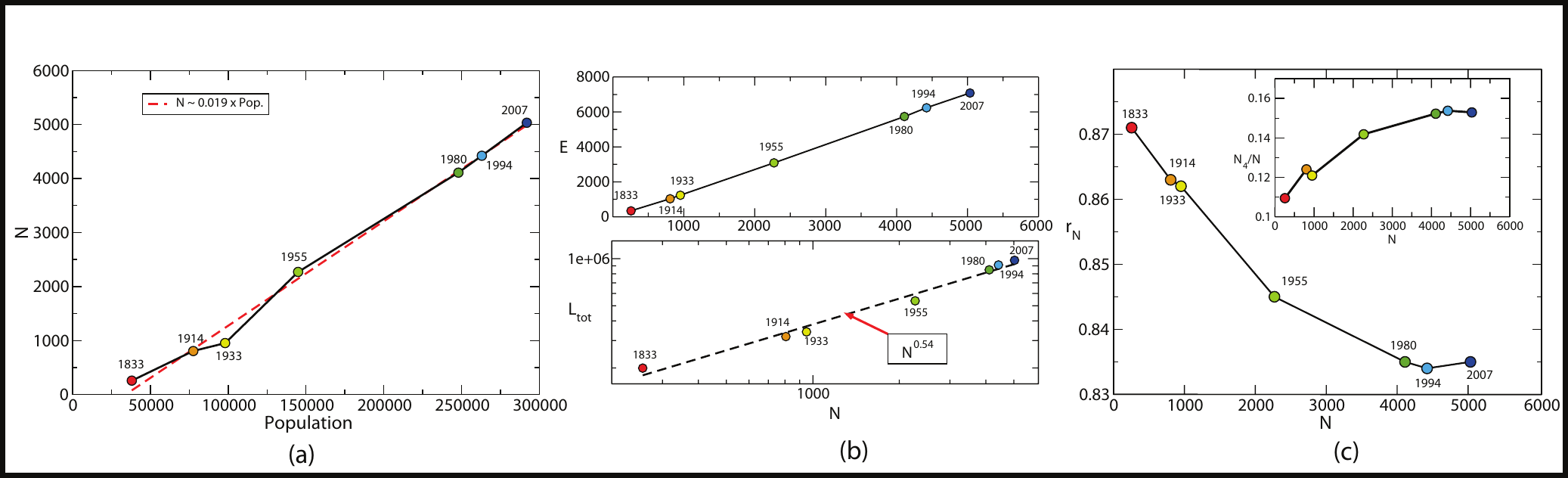}
\caption{{\bf (a)} Number of nodes $N$ versus total population
  (continuous line with circles) and its linear fit (red dashed line).
  \textbf{(b)} Total number of links $E$ and total network length
  $L_{tot}$, as a function of the number of nodes $N$. The total
  network length increases as $N^{0.54}$. {\bf (c)} Value of the ratio
  $r_N$ between the number of nodes with degree $k=1$ and $k=3$, and
  the total number of nodes. In the inset we report the percentage of
  nodes having degree $k=4$ as a function of $N$. Notice that the
  relative abundance of four-ways crossings increases by $5\%$ in two
  centuries.}
\label{fig:basic_pro}
\end{figure*}

The Groane
area is basically characterised by an uninterrupted growth in the period
under consideration and displays remarkable modifications of the road
system. As shown in Fig.~\ref{fig:panel1}c, in less than two
centuries, the total number of nodes $N$ has increased by a factor of
twenty, from the original $255$ nodes present at $t=1$ (year $1833$)
to more than $5000$ nodes at $t=7$ (year $2007$). However, the rate of
growth is not constant over time: it is slow from 1833 to 1933, fast
from 1933 to 1980, and slow again from 1980 to 2007.  These different
growth rates are the signature of distinct phases of the urbanisation
process, and are strongly related to the growth of the population. We
indeed show that the number of nodes $N$ is a linear function of the
number of people living in the Groane area (see
Fig.~\ref{fig:basic_pro}a) or, in other words, that the average number
of people per road intersection in the area remains constant over
time.

These results highlight the peculiarity of an area where the
population remains almost uniformly distributed in space, and no
single urban center stands out over the others. In order to remove the
demographic component and focus solely on the road network evolution,
we adopt the number of nodes $N$ as the natural internal clock of the
system, and we study the change of the network properties as a
function of $N$.  The number of links $E$ grows almost linearly with
$N$ (Fig.~\ref{fig:basic_pro}b, top), showing that the average node
degree $\langle k\rangle (t)=2E(t)/N(t)$ is roughly constant over
time, except for a slight increase from $\langle k\rangle\simeq 2.57$
to $2.8$ when going from 1914 to 1980.
In the bottom panel of Fig.~\ref{fig:basic_pro}b we report the total
network length $L_{tot}(t)=\sum_{e\in E_t}\ell (e)$ (where $e$ denotes
a network edge, and $\ell(e)$ its length) which increases over time as
\begin{equation}
L_{tot}\sim N^{\gamma}
\end{equation}
where $\gamma\simeq 0.54$. Consequently, the average length of links
decreases as $N^{\gamma-1}$. This result
is consistent with the evolution of almost regular two-dimensional
lattices with a peaked link length
distribution~\cite{Barthelemy:2008,Barthelemy:2011}. Indeed, if we
consider a uniform distribution of nodes with density $\rho$ on a
two-dimensional surface, the typical link length will be of order
$\ell'\sim 1/\sqrt{\rho}$, which implies that the total length grows
as $L_{tot}\sim E\ell'\sim \sqrt{N}$, thus giving a value
$\gamma=1/2$.

Additional information on the structure of the road network can be
obtained by looking at the quantity \cite{Courtat:2011} 
\begin{equation}
r_N=\frac{N_{1}+N_{3}}{\sum_{k\neq 2}N_{k}}
\end{equation}
where $N_{k}$ denotes the number of nodes of degree $k$ (notice that
we do not take into account nodes having $k=2$ in the sum, since these
are not usually considered as proper junctions). This quantity $r_N$
measures the relative abundance of dead ends (corresponding to
$N_{1}$) or T-shaped intersections ($N_{3}$), so that a small value of
$r_N$ indicates a dominance of $k=4$ junctions and reveals the
presence of a large amount of grid-like patterns. Conversely, the
value of $r_N$ is closer to $1$ if the network has numerous T-shaped
crossings and dead ends. The plot of $r_N$ versus $N$
(Fig.~\ref{fig:basic_pro}c) displays a steady decrease from $r_N\simeq
0.87$ at year $1833$ to $r_N\simeq 0.835$ at year $2007$, consistent
with an increase of the percentage of four-ways crossings from
$N_4/N\simeq 11\%$ at year $1833$ to $N_4/N\simeq 15.5\%$ at year
$2007$. Recent studies have shown that the abundance of T-shaped
crossings seems to be typical of self-organised or `organic' urban
networks, like those of Venice or Cairo~\cite{Cardillo:2006}, while
the grid-like layout is particular to cities whose shapes are the
result of large-scale top-down planning efforts~\cite{Courtat:2011},
like Barcelona or New York. Consequently, one would conclude that what
we are observing here is the evolution from an initial self-organised
system to a rationally planned urban network. However, the Groane area
at $t=1833$ was actually a fully rural, not urban-organic network, and
the successive evolution up to year $2007$ has never witnessed any
large-scale planning whatsoever. We therefore interpret the result of
Fig.~\ref{fig:basic_pro}c as the signature of an evolution from a
`pre-urban' condition, with the dominance of dead-ends and
$3-$junctions typical of rural centres in the very early stages of
growth (i.e. still constrained by the radial convergence of major
roads), to an increasingly mature `urban' state, in which the network
expands on a relatively unconstrained land, further away from the
dense radial system of original centres. At that later urban stage the
dominance of the $4-$junctions grid-like pattern does not result from
large-scale planning, but from a piecemeal urbanisation, made of a
multitude of small-scale, plot-based, scarcely coordinated
developments in time, i.e. a substantially self-organised territorial
order.

\subsection*{Evolution of cells: towards homogenisation.} 
\begin{figure*}
  \centering
  \includegraphics[width=7in]{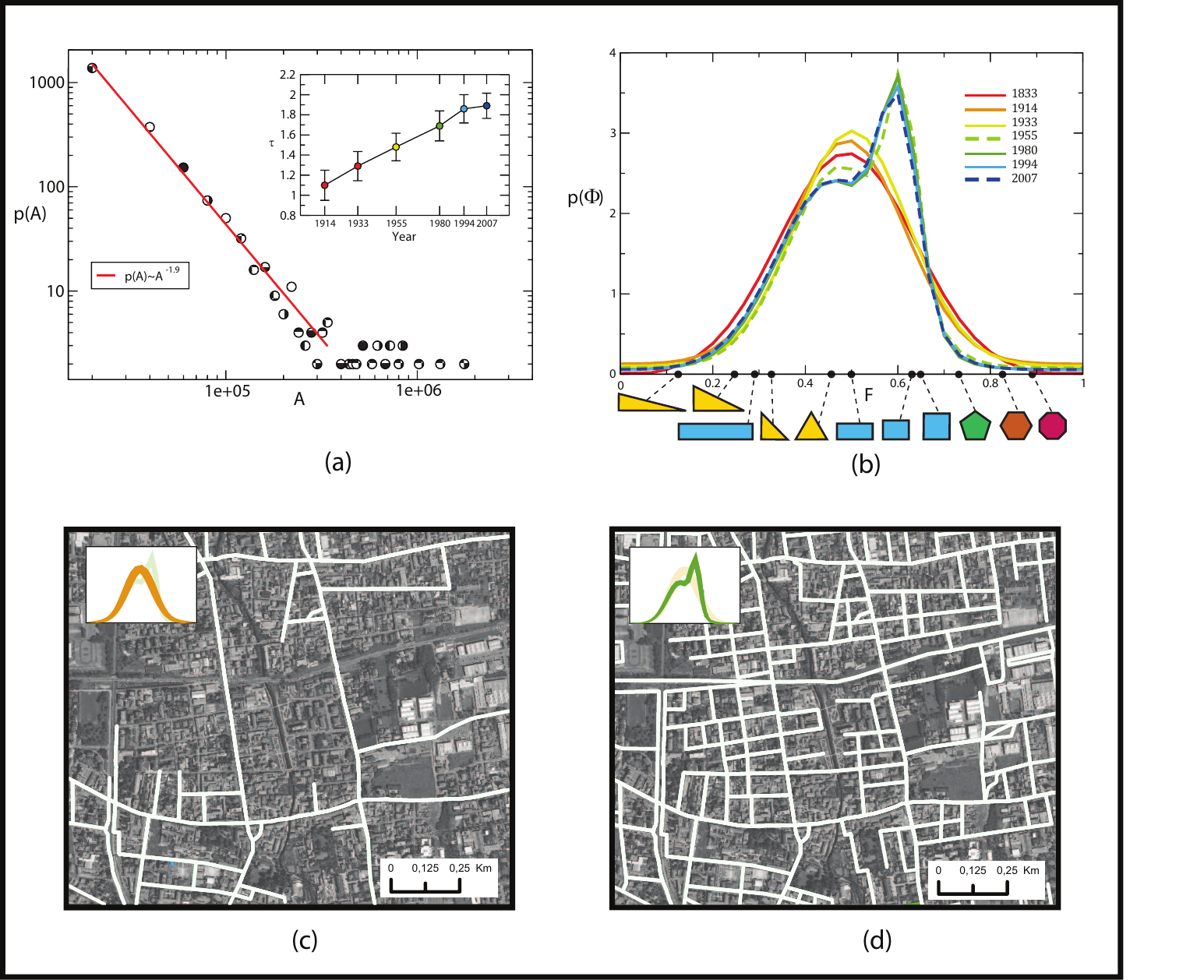}
  \caption{ {\bf (a)} The size distribution of cell areas at $t=2007$
    can be fitted with a power-law $p(A) \sim A^{-\tau}$, with an
    exponent $\tau\simeq 1.9$. The inset shows the value of $\tau$ at
    different times. {\bf (b)} Distribution of cell shapes at
    different times, as quantified by the shape factor $\Phi$. The
    shape factor of different polygons is reported at the bottom axis
    for comparison. {\bf (c)} and {\bf (d)} Maps showing the cell
    shapes (white lines) for the network as it is before 1955 (left
    panel) and as it is after 1955 (right panel). We see on the left
    panel that we have predominantly triangles and rectangles, while
    on the right panel we can observe a predominance of rectangles
    with sides of almost the same length.}
\label{fig:cells2}
\end{figure*}

Road networks are
planar graphs consisting of a series of land cells surrounded by
street segments. The statistics on the area and the shape of cells can
be used to distinguish regular lattices from heterogeneous
patterns. In particular, it has been recently observed that for
Dresden (Germany)\cite{Lammer:2006} and for a simple model of road
networks \cite{Barthelemy:2008}, the cell area distribution $P(A)$ is a
power law
\begin{equation}
P(A)\sim A^{-\tau}
\end{equation}
with an exponent $\tau$ very close to the value $1.9$.  This value of
$\tau$ can be explained in terms of a lattice constructed on a set of
nodes with density fluctuations~\cite{Barthelemy:2011}. In Fig.~\ref{fig:cells2}a we report the
distribution of the cell size in the Groane area at $t=2007$, which is
indeed a power law with the same exponent $\tau = 1.9 \pm
0.1$. However, the exponent changes in time, as reported in the inset:
it takes a value $\tau \simeq 1.2$ at year $1833$ and converges
towards $\tau \simeq 1.9$ as the network grows. Because a larger
exponent indicates a higher homogeneity of cell areas, we are thus
witnessing here a process of homogenisation of the size of cells.
Accordingly, the relative dispersion $\delta_A=\sigma(A)/\mu(A)$ of
cell areas (where $\mu(A)$ and $\sigma(A)$ are the average and the
standard deviation of $A$, respectively) decreases from $0.5$ at year
$1833$ to $0.26$ at year $2007$, indicating that the variance of the
distribution becomes smaller as $N$ increases.

The diversity of the cell shapes can be quantitatively characterised
by the so-called shape factor $\Phi=A/(\pi D^2/4)$, defined as the
ratio between the area $A$ of a cell and the area of the circle of
diameter $D$ circumscribed to the cell~\cite{Haggett:1969}. The value
of the shape factor is in general higher for regular convex polygons,
and tends to $1$ when the number of sides in the polygon
increases. The distributions $P(\Phi)$ reported in
Fig.~\ref{fig:cells2}b clearly reveal the existence of two different
regimes: before $1933$ the distributions are single-peaked and well
approximated by a single Gaussian function with an average of about
$0.5$ and a standard deviation of $0.25$. Conversely, after $1955$ the
distributions of the shape factor display two peaks and can be fitted
by the sum of two Gaussian functions.  The first peak coincides
roughly with the one observed before $1933$, while the second peak,
centred at $0.62$, signals the appearance after $1955$ of an important
fraction of regular shapes, such as rectangles with sides of similar
lengths. In Fig.~\ref{fig:cells2}c-d we show the cell shapes at
different times (before and after 1$1955$) which visually confirm the
findings of Fig.~\ref{fig:cells2}. The decrease in the relative
dispersion $\delta_A$ and the increase in the fraction of regular
shapes suggest that the network undergoes an evolution towards
homogenisation. This appears to result from a combined process that
exhibits two clear patterns: \textit{i)} the fragmentation of larger
cells of natural land into smaller ones, then heading towards
urbanisation by medium-large manufactures or services, and
\textit{ii)} the mostly residential urbanisation of peri-urban natural
land in successive rings around the historical urban centres.  In
particular, this latter pattern explains the emergence of more
regularly shaped cells after the Second World War. As urbanization
took place increasingly further from historical main roads,
residential blocks became less constrained by the triangular shape
defined by those roads and, as a result, increasingly followed a
regular rectangular shape. This regular shape is in fact the most
efficient way of subdividing land for urbanization, reflecting the
inner organization of the block into equally regular rectangular
plots. This grid-like layout tends to be applied extensively when
local constraints, like main roads converging into a village, do not
force the development along different patterns (for example T-shaped
junction dominated) \cite{Ben-Joseph:2005}.

%--------------------------------------------
\subsection*{Properties of new links: elementary processes of
  urbanisation.}  
\begin{figure}
\centering 
\includegraphics[width=3in]{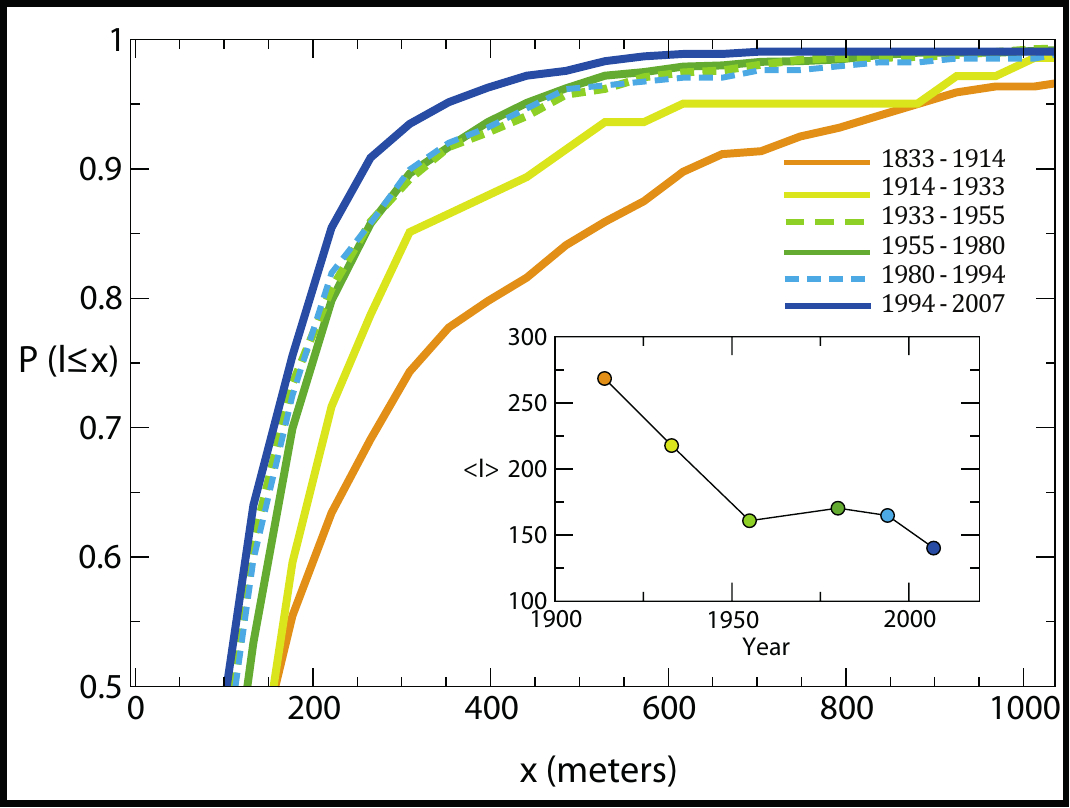}   
  \caption{Cumulative distributions of the length of links added at
    different times. In the inset we report the average length
    $\langle\ell\rangle$ of new links.}
  \label{fig:new_links}
\end{figure}

Road networks grow by the addition of new streets
(links) and new junctions (nodes). We focus here on the properties of
new links by looking at their length and centrality. In
Fig.~\ref{fig:new_links} we show the cumulative distribution of the
length of new links according to the time-section in which they
appeared first. The inset shows that the average length of new links
steadily decreases over time, as expected from the general
considerations of land fragmentation reported above. More precisely,
we consider, at each time $t$, the length value $\ell_{90\%}(t)$ such
that $90\%$ of new links at time $t$ are shorter than $\ell_{90\%}(t)$
(i.e. such that $P(\ell \le \ell_{90\%})=0.9$). We notice that the
value $\ell_{90\%}$ decreases in the period 1833-1933 from $625$
meters down to $325$ meters, while no sensible variation is observed
from 1933 to 1994, even if the network keeps growing. In the last
period, i.e. from 1994 to 2007, we observe another decrease of
$\ell_{90\%}(t)$ from $325$ to $225$ meters. In addition, the relative
dispersion of the length of the new links is almost constant and of
order one, and the distribution does not vary too much after
1955. This phasing fits well with the historical development outlined
in the introduction where during the rural and early-urban phase up to
the second world war we observe the passage to an urban state that is
then maintained along the core urbanisation age in the
urban-industrial period until the 1980s, followed by a different
post-industrial, metropolitan regime.

\begin{figure*}
  \centering
  \includegraphics [width=7in] {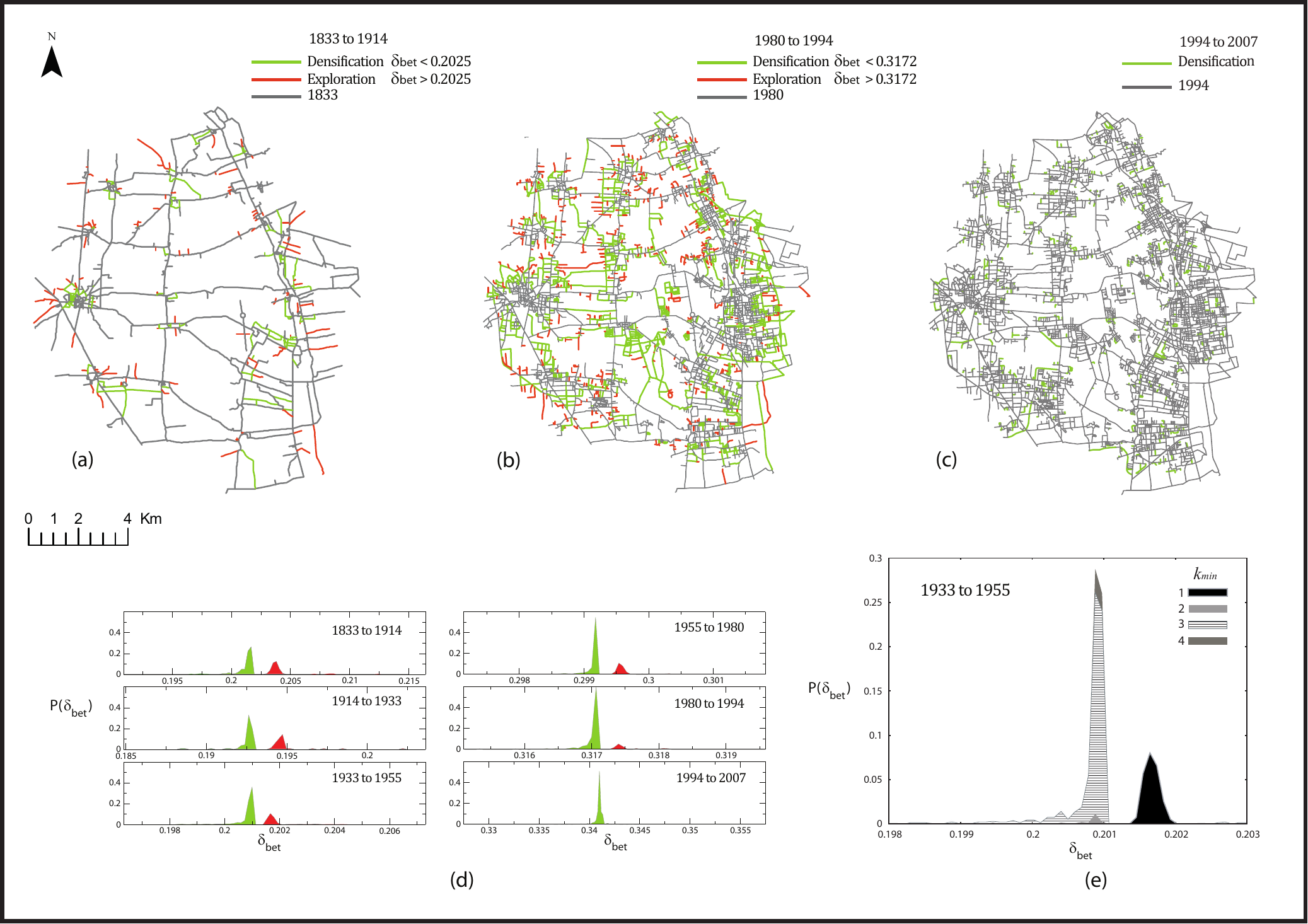}	
  \caption{The two phases of densification (green) and exploration
    (red), illustrated for the networks at year 1914 {\bf (a)}, 1994
    {\bf (b)} and 2007 \textbf{(c)}. Panel \textbf{(d)} shows the
    probability distribution of the BC impact $\delta_b(e)$ for the
    different time snapshots. The red peak corresponds to exploration,
    and the green peak to densification. Notice that the red peak
    becomes smaller and smaller with time, and completely disappears
    in the last snapshot. \textbf{(e)} Stacked chart showing the
    proportion of new links with a certain value of $k_{min}$ for each
    value of $\delta_{b}$ in the network at $t=1955$. The peak
    corresponding to exploration with a high BC impact is entirely due
    to links having $k_{min}=1$.}
  \label{fig:contribution}
\end{figure*}

The nature of the growth process can be quantitatively characterised
by looking at the centrality of streets. Among the various centrality
indices available for spatial networks, we use here the betweenness
centrality (BC) \cite{Freeman:1977,Crucitti:2006,Porta:2006}, which is
one of the measures of centrality commonly adopted to quantify the
importance of a node or a link in a graph. Given the graph $G_t\equiv
G(V_t,E_t)$ at time $t$, the BC of a link $e$ is defined as 
\begin{equation}
  b(e) = \sum_{i\in V} \sum_{\stackrel{j \in V} {j \neq i}} 
\frac{\sigma_{ij}(e)}{\sigma_{ij}}
\end{equation}
where $\sigma_{ij}$ is the number of shortest paths from node $i$ to
node $j$, while $\sigma_{ij}(e)$ is the number of such shortest paths
which contain the link $e$. The quantity $b(e)$ essentially measures
the number of times a link is used in the shortest paths connecting
any pair of nodes in the network, and is thus a measure of the
contribution of a link in the organisation of flows in the network. In
order to evaluate the impact of a new link on the overall distribution
of the betweenness centrality in the graph at time $t$, we first
compute the average betweenness centrality of all the links of $G_t$
as:
\begin{equation}
\overline{b}(G_t) = \frac{1}{(N(t) - 1) (N(t) - 2)} \sum_{e\in E_t} b(e)
\end{equation}
where $b(e)$ is the betweenness centrality of the edge $e$ in the
graph $G_t$. Then, for each link $e^* \in \Delta E_t$, i.e. for each
newly added link in the time window $]t-1,t]$ we consider the new
graph obtained by removing the link $e^*$ from $G_t$ and we denote
this graph as $G_t \setminus \{ e^* \}$.  We compute again the average
edge betweenness centrality, this time for the graph $G_t \setminus \{
e^* \}$. Finally, the impact $\delta_b(e^*)$ of edge $e^*$ on the
betweenness centrality of the network at time $t$ is defined as
\begin{equation}
\delta_b(e^*) = \frac { \left[\overline{b}(G_t) - \overline{b}( G_t
    \setminus \{ e^* \} ) \right] }{ \overline{b}(G_t)}
\end{equation}
The BC impact is thus the relative variation of the graph average
betweenness due to the removal of the link $e^*$. We have measured
this quantity for the different times-section and we report the
results in Fig.~\ref{fig:contribution}. Remarkably, the distribution
of $\delta_b(e)$ displays two well-separated peaks
(Fig.~\ref{fig:contribution}d). The importance of the first peak tends
to increase in time while the second peak decreases, until they mostly
merge into one single peak in the last time-section (1994-2007). In
order to understand the nature and the evolution of the two peaks, we
focus on the geographical location of new links with different BC
impact. 

In Fig.~\ref{fig:contribution}a-c we report, respectively, the map of
the network at year $1914$, year $1994$ and year $2007$. We have coloured in green
the links whose centrality impact belongs to the left peak, and in red
the links whose $\delta_b$ corresponds to the right peak. The new
links are divided into two classes: green links (small $\delta_b$,
left peak) tend to bridge already existing streets, while red links
(large $\delta_b$, right peak) are usually dead-ends edges branching
out of existing links, generating a new crossing and splitting the
original link into two road segments. In order to better characterise
the two classes of edges we compute $k_{min}(e)$, i.e. the minimum of
the degrees of both endpoints of a new link $e$, and we then consider,
for each value of $\delta_b$, the relative abundance of new links for
which $k_{min}$ is equal to $1,2,3,4$, respectively. We plot the
results in a gray--scale stacked chart (see
Fig.~\ref{fig:contribution}e) for the network at year $1955$ (the
plots for the other networks are similar). The analysis of $k_{min}$ confirms that the links belonging to
different peaks have distinct qualitative features: links in the right
peak have $k_{min}=1$, while links in the left peak have $k_{min}\geq
2$.  The distribution of BC impact thus suggests that the evolution of
the road network is essentially characterised by two distinct,
concurrent processes: one of \textit{densification} (green links, left
peak, i.e. lower impact on centrality, $k_{min}\geq 2$) which is
responsible for the increase of the local density of the urban
texture, and one of \textit{exploration} (red links, right peak,
higher impact on centrality, $k_{min}=1$) which corresponds to the
expansion of the network towards previously non-urbanised areas.  At a
closer sight, these two patterns tend in many cases to appear in a
temporal sequence, the former acting in preparation for the latter,
i.e.  exploration being the first phase of a cycle of urbanisation
then completed by a second phase of densification.  Obviously, since
the amount of available land decreases over time, at earlier
time--sections (such as in the year $1914$) the fraction of
exploration is higher, while in the $80$'s it becomes smaller until it
almost disappears in 2007. The only remaining peak mostly corresponds
to densification with new links having $k_{min}\geq 2$.

\begin{figure*}
  \centering
  \includegraphics [width=7in] {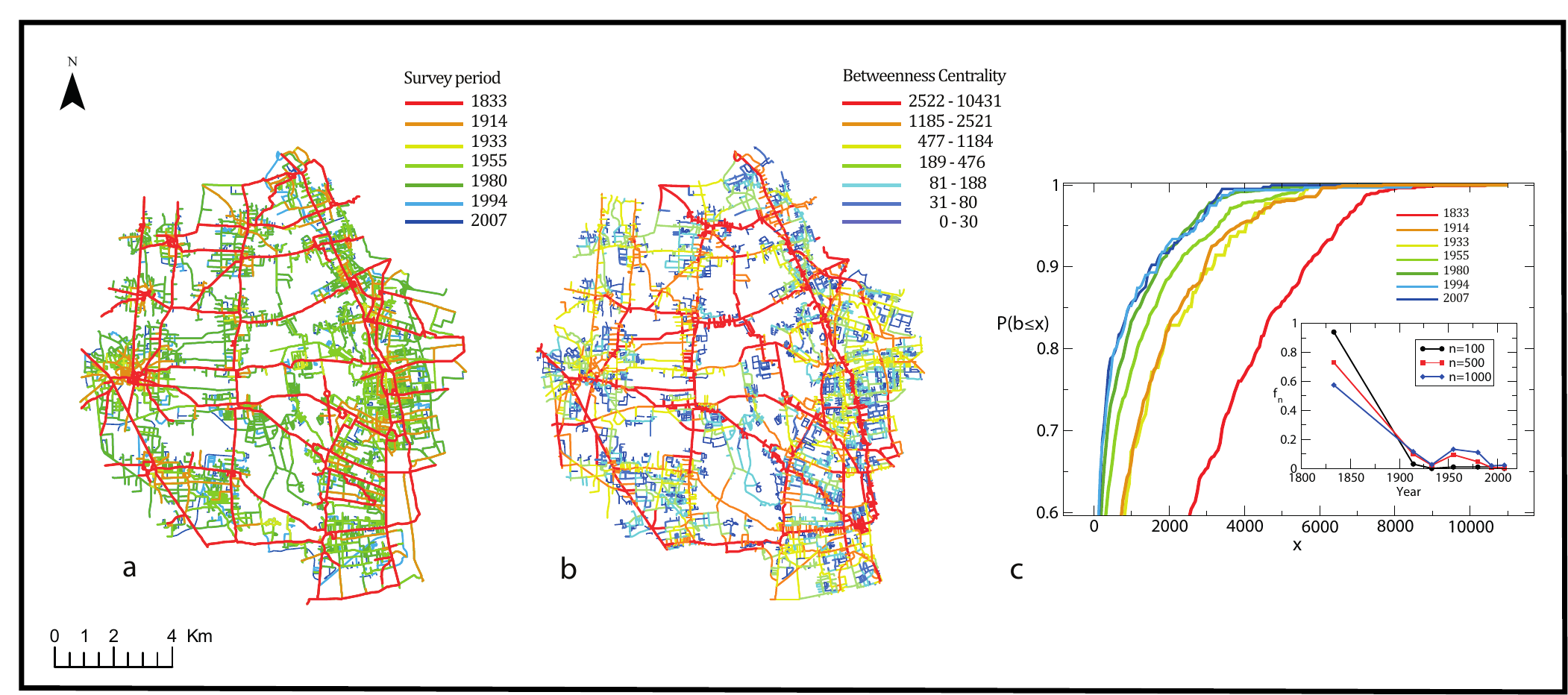}	
  \caption{Color maps indicating {\bf (a)} the time of creation of each link
    and {\bf (b)} its value of betweenness centrality
    (BC) at year 2007. \textbf{(c)} The cumulative
    distribution of BC of links added at different times. The inset
    reports the percentage of edges added at a certain time which
    are ranked in the top $n$ positions according to the BC. Different
    curves correspond to $n=100, 500, 1000$.}
\label{fig:bet_panel}
\end{figure*}

We finally focus on another aspect of the evolution of street
centrality, namely the relation between the age of a street and its
centrality.  In Fig.~\ref{fig:bet_panel}a we display the links of the
network in $2007$ with a color code depending on their age, and in
Fig.~\ref{fig:bet_panel}b we report their BC.  A simple visual
inspection shows that highly central links usually are also the oldest
ones. In particular, the links constructed before $1833$ have a much
higher centrality than those added at later time--sections. More
precisely, the seven curves in panel Fig.~\ref{fig:bet_panel}c report
the cumulative distribution of the BC computed on the network in
$2007$ for the links added at the different times. This is defined as
the probability $P(b \le x)$ that a link, appearing at a certain
time-section, has a value of betweenness centrality $b$ smaller than
or equal to $x$ in the final network in $2007$.  We notice that
the historical structure of the oldest roads mostly coincide with
the highly central links at year $2007$. In particular, the inset of
Fig.~\ref{fig:bet_panel}c indicates that more than $90\%$ of the $100$
most central links in 2007 (and almost $60\%$ of the top $1000$) were
already present in 1833. This result reveals that a "backbone" of
highly--central routes, that have framed the Groane area in the rural
economy of the pre-industrial period, has been driving the development
of the area across two centuries of industrialisation, urbanisation
and de-industrialisation up to present days, without any major
modification.

Urban morphologists have long observed that streets tend to persist in
time much longer than other urban elements like buildings or land-uses
\cite{Mouton:1989}. Our results confirm that at another level, that of
the topology of the street network, the persistence of roads over time
appears to be due to their multifaceted impact on the urban form at
different scales by deeply informing factors like accessibility to
space and resources, land ownership and land values, each of these
involving different social actors and interests.

%%%%%%%%%%%%%%%%%%%%%%%%%%%%%%%%%%%%%%%%%%%%%%%%%%%%%%
\section*{Discussion}

In this paper we have studied the evolution, over almost two
centuries, of the street network in a large large area close to Milan
(Italy). Such an area is of interest for urban studies because it
displays an important process of urbanisation. Urbanisation is
reflected in the growth of the road network and displays different
speeds at different times, with a fast increase occurring between 1933
and 1994. Our results reveal a quantitative signature of urbanisation
on the evolution of the shape and size of land cells, which become
more homogeneously distributed and square-shaped. Simultaneously, we
observe a general trend towards a larger number of 4-way junctions, as
opposed to an earlier structure of dead ends and 3-way
junctions. These structural transformations appear to be the result of
the interplay between two concurrent dynamics, namely densification
and exploration. While exploration is typical of the earliest
historical periods of urbanisation, densification predominates in the
latest.

We were also able to quantitatively characterise the stability of the
structure of most central streets over time. For instance, the most
central streets in the network in $2007$ largely coincide with
the oldest ones. Central roads appear to therefore constitute a robust
spatial backbone which remains stable over time, and characterises the
evolution of the road system as a continuous expansion and
reinforcement of pre-existing structures rather than as a sharp switch
towards radically new configurations. The kind of evolution that we
are witnessing in the Groane area is therefore certainly massive in
overall quantitative terms, reflecting unparalleled changes in the
pace and nature of economic, technological and social order during the
transition from a mainly agricultural to a post-industrial age;
however, it is still an organic form of evolution that builds at each
step on the achievements of the previous, confirming in the long term
a resilient structure that tends to persist in time all over the
urbanisation process. These findings quantitatively support the
hypothesis that spatial systems undergoing fine-grained forms of
evolution tend to exhibit simple local dynamics of change,
continuously expanding upon pre-existing structures, a morphogenetic
behaviour that they appear to share with many living organisms and
other natural systems~\cite{Alexander:2003}.

The results of this research, including the existence of the two
main dynamics of exploration and densification, together with the
persistence over time of a structural backbone made of highly central
streets, cannot be extended to the generality of urbanisation
processes unless supported by further investigation of cases in
different geographical and economic positions. An important point in
our view is that this study provides quantitative results which can be
confirmed (or falsified), a critical aspect in Science.

\section*{Methods}

\subsection*{Temporal Network Data.}  The area under study covers
$125\text{km}^2$ and includes $29$ urban centers within $14$
municipalities that have developed along two main radial paths,
connecting Milan to Como and Milan to Varese. We sampled seven street
networks primal graphs at different times, where the street junctions
are represented as nodes and the roads (or streets) are the links of
the networks. We used a mixed procedure of ArchMap tool extensions and
python scripts operating over a geo-database.  First, the street
network has been drawn in vectorial format on the basis of a
collection of historical areal images and historical maps imported in
the ArcGIS10 platform. Sources used for each temporal steps are
reported in Table~$1$. The images have been imported as a .geotif
extension, and all the layers has been projected using the coordinate
system Monte Mario (Rome), Italy zone and Transverse Mercator
projection. The first geo-referencing (alignment of different images)
has been performed on the basis of historical buildings and landmarks
such as paths and roads. The redraw operation was done using an ad-hoc
python script for creating the node layers representing the street
junctions. The second geo-referencing operation has been performed
over the street junctions using the “spatial analyst tool” with the
“spatial join” utilities provided in ArchMap10.  Subsequently, for all
time layers, using an ad-hoc python script we produced the weighted
adjacency lists that has been used for the network analysis, the
weight correspond to the real length of the street. We denote by $G_t
\equiv G(V_t, E_t)$ the obtained primal graph at time $t$, where $V_t$
and $E_t$ are respectively the set of nodes and links at time $t$.
The number of nodes at time $t$ is then $N(t)= |V_t|$ and the number
of links is $E(t) = |E_t|$. Using common definitions, we thus have
$V_t = V_{t-1} \cup \Delta V_t$ and $E_t = E_{t-1} \cup \Delta E_t$,
where $\Delta V_t$ and $\Delta E_t$ are respectively the new street
junctions and the new streets added in time $]t-1,t]$ to the network
    existing at time $t-1$.

%%%%%%%%%%%%%%%%%%%%%%%%%%%%%%%% references

\begin{acknowledgments} MB thanks R. Morris for useful comments and
the Morphocity group (P. Bonnin, P. Bordin, C.-N. Douady, S. Douady,
J.-P. Frey, P. Vincent) for stimulating discussions. ES and SP are
grateful to the Laboratory of Territorial Planning at DIAP,
Polytechnic of Milan (S. Serini, A. Benedetti L. Terlizzi, G. Graj and
Prof. P. L. Paolillo) for providing cartographic material and
vectorial street layers. They also thank A. Venerandi for his help in
the first phase of this study.
\end{acknowledgments}

\section*{Author contributions}

All the authors have equally contributed to the design of the
experiment, to the analysis and interpretation of the results and to
the preparation of the manuscript.

\section*{Additional Information}

{\bf Competing financial interests}\\ The authors declare no
competing financial interests.

\end{document}